\documentclass[11pt]{article}
\usepackage{amsmath,amssymb,amsfonts,graphicx,cite,soul}
\usepackage{epsfig}
\usepackage{axodraw}
\usepackage{slashed}

\def\al{\alpha}

\newcommand{\eqFeyn}[1]{%
\begin{array}{c} #1 \end{array}
}
\newcommand{\be}{\beta}
\newcommand{\mD}{m_D}
\newcommand{\mM}{m_M}

\newcommand{\mR}{m_{\tilde R}}

\newcommand{\Anu}{A_\nu}

\newcommand{\Bnu}{B_\nu}
\newcommand{\mnu}{m_\nu}
\newcommand{\Ynu}{Y_\nu}
\def\hSi{\hat{\Sigma}}
\def\Si{\Sigma}

\newcommand{\DRbar}{\ensuremath{\overline{\mathrm{DR}}}}
\newcommand{\mDRbar}{\ensuremath{\mathrm{m}\overline{\mathrm{DR}}}}

\newcommand{\cO}{{\cal O}}

\newcommand{\wz}{\sqrt{2}}
\newcommand{\edz}{\frac{1}{2}}

\newcommand{\fa}{{\em FeynArts}}
\newcommand{\fc}{{\em FormCalc}}
\newcommand{\fh}{{\tt FeynHiggs}}

\newcommand{\MW}{M_W}
\newcommand{\MZ}{M_Z}

\newcommand{\MA}{M_A}

\newcommand{\Mh}{M_h}
\newcommand{\MH}{M_H}

\newcommand{\Snu}{\tilde \nu}

\newcommand{\tsf}{\theta\kern-.20em_{\tilde{f}}}
\newcommand{\tsfp}{\theta\kern-.20em_{\tilde{f}\prime}}
\newcommand{\tsq}{\theta\kern-.15em_{\tilde{q}}}

\newcommand{\se}[1]{\Sigma_{#1}}
\newcommand{\ser}[1]{\hat{\Sigma}_{#1}}

\newcommand{\KL}{\left(}
\newcommand{\KR}{\right)}
\newcommand{\KKL}{\left[}
\newcommand{\KKR}{\right]}

\newcommand{\VL}{\left( \begin{array}{c}}
\newcommand{\VR}{\end{array} \right)}
\newcommand{\ML}{\left( \begin{array}{cc}}
\newcommand{\MLd}{\left( \begin{array}{ccc}}
\newcommand{\MLv}{\left( \begin{array}{cccc}}
\newcommand{\MR}{\end{array} \right)}

\newcommand{\tb}{\tan \beta}

\newcommand{\CTb}{\cot \beta}

\newcommand{\CZb}{\cos 2\beta}

\newcommand{\gev}{\,\, \mathrm{GeV}}
\newcommand{\mev}{\,\, \mathrm{MeV}}

\newcommand{\non}{\nonumber}
\newcommand{\id}{{\rm 1\kern-.12em
\rule{0.3pt}{1.5ex}\raisebox{0.0ex}{\rule{0.1em}{0.3pt}}}}

\newcommand{\lsim}

\newcommand{\gsim}

\newcommand{\tadH}{T_H}
\newcommand{\tadh}{T_h}

\newcommand{\tanb}{\tan \beta\,}

\newcommand{\dmhsq}{\delta m_h^2}

\newcommand{\dZ}[1]{\delta Z_{#1}}

\def\al{\alpha}

\begin{document}
\begin{flushright}
IFT-UAM/CSIC-12-06
\end{flushright}

\vspace{0.2cm}

{\large\sc {\bf Heavy Majorana neutrino effects on MSSM-Mh}}

\vspace{0.6cm}

{\sc
M.J.~Herrero$^{1}$
\footnote{Talk given at RADCOR2011, September 26-30, 
2011, Mamallapuram, India.}%
, S.~Heinemeyer$^{2}$, S.~Pe\~naranda$^{3}$
\footnote{Present address: Departament de F{\'\i}sica Fonamental,
Universitat de Barcelona, Spain}%

~and A.M.~Rodr\'iguez-S\'anchez$^{1}$%
}

\vspace*{.7cm}

{\sl
$^1$Instituto de F\'isica Te\'orica, UAM/CSIC, Madrid, Spain

\vspace*{0.1cm}

$^2$Instituto de F\'isica de Cantabria, IFCA-CSIC, Santander, Spain

\vspace*{0.1cm}

$^3$Departamento de F\'isica Te\'orica, Universidad de Zaragoza, Spain

\vspace*{0.1cm}

E-mails: maria.herrero@uam.es, sven.heinemeyer@cern.ch, 
siannah@unizar.es, anam.rodriguez@uam.es

}

\vspace*{0.1cm}

\begin{abstract}
\noindent
We study the effects of heavy Majorana neutrinos on the Higgs sector 
of the MSSM via radiative corrections. We work within the SUSY context where 
the MSSM particle content is 
enlarged with right handed neutrinos and their corresponding SUSY partners, 
the sneutrinos, and where compatibility with neutrino data is required. 
We compute the one-loop 
corrections to the mass of the lightest MSSM CP-even neutral Higgs boson from 
Majorana neutrinos and their SUSY partners and assume a seesaw mechanism 
of type I for neutrino mass generation. A negative and sizeable 
Higgs mass correction of up to -5 GeV is found for a heavy Majorana mass of 
up to $10^{15}$ GeV. This negative correction can grow up to several tens 
of GeV if the soft SUSY breaking mass associated to their sneutrino partners 
is simmilarly heavy as the Majorana mass.
\end{abstract}

%%%%%%%%%%%%%%%%%%%%%%%%%%%%%%%%%%%
\section*{Introduction}
The current %impressive
 experimental data %\cite{pdg} 
on neutrino mass differences and 
neutrino mixing angles clearly indicate new physics beyond
the so far successful Standard Model of Particle Physics (SM). In particular, 
neutrino oscillations imply that at
least two generations of neutrinos must be massive. Therefore, one needs to extend the SM to 
incorporate neutrino
mass terms.

 We assume here the simplest version of a SUSY extension of the SM, the well known
 Minimal Supersymmetric Standard Model (MSSM), extended by right-handed Majorana neutrinos and
 their SUSY partners, and 
where the seesaw mechanism of type I~\cite{seesaw:I} is implemented to generate the small 
neutrino masses. For simplicity, we focus here in the one generation case which already provides
interesting results and allows for a clearer illustration of the most relevant behaviour of the
radiative corrections with the Majorana mass scale. The three generation
case will be postponed for a later work.

%One interesting property of introducing Majorana neutrinos is that Lepton Number is no longer conserved, and leptogenesis
%becomes a viable mechanism for explaining the mistery of the Baryon Asimmetry of the Universe (BAU) 
%when these Majorana neutrinos are heavy enough. %($\mM \geq 10^{12}$ GeV).

On the other hand, it is well known that heavy Majorana neutrinos, with $m_M \sim 10^{13}-10^{15}$ GeV, 
 induce large LFV rates~\cite{LFV}, due to their potentially large Yukawa
 couplings
to the Higgs sector. For the same reason, radiative corrections to Higgs boson masses due to such heavy Majorana neutrinos
could also be relevant. %as has already been pointed out  in particular scenarios in previous works~\cite{Cao:2004hs,Kang,Dedes:2007ef}.
Consequently, our study has been  focused on the 
radiative corrections to the lightest MSSM CP-even $h$ boson mass, $\Mh$,
due to the one-loop contributions from the neutrino/sneutrino sector 
within the MSSM-seesaw framework.

 In the following we briefly review the main relevant aspects of the calculation of the mass corrections and the
numerical results. 
For further details we address the reader to the full version of our work~\cite{Heinemeyer:2010eg},
 where also an extensive list with references to related works can be found.
 %Furthermore, we have compared our results with ~\cite{Cao:2004hs}, where they work in an split SUSY scenario
%and they  obtain large corrections when the soft-SUSY-breaking mass associated to the 
%right handed neutrino, $\mR$, was chosen to be very large, of the 
%order of the Majorana scale $\mM$. However, in comparison with them we don't neglect the mixing in the sneutrino sector
%and we analize also the standard SUSY scenario with soft masses of O(TeV) and we get relevant corrections also in this case.
%%%%%%%%%%%%%%%%%%%%%%%%%%%%%%%%%%%%%%%%%%%%%%%%%%%%%%%%%%%%%%%%%%%%%%%%%%%%%%%%%%%%%%%%%%%%%%%%%%%%%%
\section*{The basics of the MSSM-seesaw model}
%\subsection*{The neutrino/sneutrino sector}
\label{sec:nusnu}

The MSSM-seesaw model with one neutrino/sneutrino generation is described
in terms of the well known MSSM superpotential plus the new
relevant terms contained in: %~\cite{Gunion:1986yn}
\begin{equation}
\label{W:Hl:def}
W\,=\,\epsilon_{ij}\left[\Ynu \hat H_2^i\, \hat L^j \hat N \,-\, 
Y_l \hat H_1^i\,\hat L^j\, \hat R  \right]\,+\,
\edz\,\hat N \,\mM\,\hat N \,,
\end{equation}
where $\mM$ is the Majorana mass and  $\hat N = (\Snu_R^*, (\nu_R)^c)$ is 
the additional superfield that contains the  right-handed 
neutrino $\nu_{R}$ and its scalar partner $\Snu_{R}$.   

There are also new relevant terms in the soft SUSY breaking potential: %~\cite{Grossman:1997is}:
%due to the additional sneutrinos $\Snu_R$ 
\begin{equation}
V^{\Snu}_{\rm soft}= m^2_{\tilde L} \Snu_L^* \Snu_L +
  m^2_{\tilde R} \Snu_R^* \Snu_R + (\Ynu \Anu H^2_2 
  \Snu_L \Snu_R^* + \mM \Bnu \Snu_R \Snu_R + {\rm h.c.})~.
\end{equation}

After electro-weak (EW) symmetry breaking, the charged lepton and 
Dirac neutrino masses
can be written as
\begin{equation}
m_l\,=\,Y_l\,\,v_1\,, \quad \quad
\mD\,=\,\Ynu\,v_2\,,
\end{equation}
where $v_i$ are the vacuum expectation values (VEVs) of the neutral Higgs
scalars, with $v_{1(2)}= \,v\,\cos (\sin) \be$ and $v=174 \gev$.

The $ 2 \times 2$ neutrino mass matrix is given in terms of $\mD$ and
$\mM$ by: 
\begin{equation}
\label{seesaw:def}
M^\nu\,=\,\left(
\begin{array}{cc}
0 & \mD \\
\mD & \mM
\end{array} \right)\,. 
\end{equation}
Diagonalization of $M^\nu$ leads to two mass
eigenstates, $n_i \,(i=1,2)$, which are Majorana fermions
with the respective  mass eigenvalues given by:
\begin{equation}
\label{nuigenvalues}
m_{\nu,\, N}  = \edz \KL \mM \mp \sqrt{\mM^2+4 \mD^2} \KR~. 
\end{equation}  
The mixing angle that defines the mass eigenstates is given by,
\begin{equation}
\label{nuangle}
\tan \theta = -\frac{\mnu}{\mD}= \frac{\mD}{m_N}~.
\end{equation}
In the seesaw limit, i.e. when $\xi \equiv \frac{\mD}{\mM} \ll 1~$:
\begin{align}
m_{\nu}= -\mD \xi + \mathcal{O}(\mD \xi^3) \simeq -\frac{\mD^2}{\mM} ~,
\quad \quad m_N    &=  \mM + \mathcal{O}(\mD \xi) \simeq \mM ~, 
\end{align}
and $\theta$ is small, therefore, 
$\nu$ is made predominantly of $\nu_L$ and its c-conjugate, $(\nu_L)^c$, 
whereas $N$
is made predominantly of $\nu_R$ and its c-conjugate, $(\nu_R)^c$.
   
Regarding the sneutrino sector, the sneutrino mass matrices for the
CP-even, ${\tilde M}_{+}$,  and the CP-odd, ${\tilde M}_{-}$,
subsectors are given respectively by % ~\cite{Grossman:1997is}: 
\begin{equation}
{\tilde M}_{\pm}^2=
\left( 
\begin{array}{cc} m_{\tilde{L}}^2 + \mD^2 + \edz \MZ^2 \cos 2 \be & 
\mD (A_{\nu}- \mu \CTb \pm \mM) \\  
\mD (A_{\nu}- \mu \CTb \pm \mM) &
m_{\tilde{R}}^2+\mD^2+\mM^2 \pm 2 \Bnu \mM \end{array} 
\right)~.
\end{equation}
The diagonalization of these two matrices, ${\tilde M}_{\pm}^2$, 
leads to four sneutrino mass eigenstates, ${\tilde n}_i \,(i=1,2,3,4)$. In the seesaw limit,
 where $\mM$ is much bigger than all the
other mass scales the corresponding sneutrino masses are given by: 
\begin{eqnarray}
 m_{{\Snu_+},{\Snu_-}}^2
&=& m_{\tilde{L}}^2 + 
\edz \MZ^2 \CZb \mp 2 \mD (A_{\nu} -\mu \CTb-\Bnu)\xi ~, \non \\
 m_{{\tilde N_+},{\tilde N_-}}^2  &=& \mM^{2} \pm 2 \Bnu \mM + \mR^2 + 2 \mD^2 ~.
\end{eqnarray}  
The mixing angles defining the sneutrinos mass eigenstates, $\theta_{\pm}$, are small in this limit and,
therefore, ${\Snu_+}$ and ${\Snu_-}$ are made predominantly of 
${\Snu_L}$ and its c-conjugate, ${\Snu_L}^*$, whereas ${\tilde N_+}$ and 
${\tilde N_-}$ are made predominantly of ${\Snu_R}$ and its c-conjugate,
 ${\Snu_R}^*$. 
 
Regarding the relevant interactions for the present work, there are pure gauge interactions 
between the
left-handed neutrinos and the $Z$~boson, 
those between the 'left-handed' sneutrinos and the Higgs bosons, and those 
between the 'left-handed' sneutrinos and the $Z$~bosons. All of these are common to the 
MSSM. In addition, in this MSSM-seesaw scenario, there are interactions 
driven by the neutrino Yukawa couplings (or equivalently $\mD$ since 
$\Ynu=(g\mD)/(\wz \MW \sin\be)$), as for instance $g_{h\nu_L \nu_R} =  -\frac{igm_D \cos\alpha}{2M_W\sin\beta}$, 
and new interactions due to the
Majorana nature driven by $\mM$ which are not present in the case of Dirac
fermions, as for instance $g'_{h\Snu_L\Snu_R}=-\frac{igm_Dm_M\cos\alpha}{2M_W\sin\beta}$. 

Finally, concerning the size of the new  parameters that have been introduced 
in this model, in addition to those of the MSSM, i.e.,
$\mM$, $\mD$, $\mR$, $\Anu$ and $\Bnu$, 
there are no significant constraints. In the literature it is 
often assumed that $\mM$ has a very large value, 
$\mM \sim \cO (10^{14-15}) \gev$, in order to
get small physical neutrino masses  $|\mnu| \sim$ 0.1 - 1 eV with 
large Yukawa couplings $ \Ynu \sim \cO(1)$. This is an interesting
possibility since it can lead to important phenomenological implications
due to the large size of the radiative corrections driven by these large
Yukawa couplings.  We have explored, however, not only these extreme values but 
the full range for $\mM$ from the
electroweak scale $\sim 10^2 \gev$ up to $\sim 10^{15} \gev$. 

\section*{The basics of the calculation}

In the Feynman diagrammatic (FD) approach the higher-order corrected 
CP-even Higgs boson masses in the MSSM, denoted here as $\Mh$ and $\MH$, 
are derived by finding the 
poles of the $(h,H)$-propagator 
matrix, which is equivalent to solving the following equation \cite{mhcMSSMlong}: 
\begin{equation}
\left[p^2 - m_{h}^2 + \hSi_{hh}(p^2) \right]
\left[p^2 - m_{H}^2 + \hSi_{HH}(p^2) \right] -
\left[\hSi_{hH}(p^2)\right]^2 = 0\,.
\label{eq:proppole}
\end{equation}
where $m_{h,H}$ are the tree level masses.
The one loop renormalized self-energies, $\hSi_{\phi\phi}(p^2)$,  in \eqref{eq:proppole} can be expressed
in terms of the bare self-energies, $\Si_{\phi\phi}(p^2)$, the field
renormalization constants $\delta Z_{\phi\phi}$  and the mass counter terms  $\delta m_{\phi}^2$, where $\phi$ stands for
$h,H$.
 For example, the lightest Higgs boson renormalized self energy reads:

\begin{equation}\label{rMSSM:renses_higgssector}
%\label{renSEhh}
\ser{hh}(p^2)  = \se{hh}(p^2) + \dZ{hh} (p^2-m_h^2) - \dmhsq,
\end{equation}

%\subsection*{Renormalization prescription}
 Regarding the renormalization prescription, we have used an on-shell renormalization scheme for $\MZ, \MW$ and $\MA$ mass counterterms 
 and $\tadh, \tadH$  tadpole counterterms. On the other hand, we have used a modified $\DRbar$ scheme  $\left(\mDRbar\right)$
 for the renormalization of the wave function 
and $\tanb$. The m$\DRbar$ scheme  is very similar to
the well known $\DRbar$ scheme but instead of subtracting the usual $\Delta= \frac{2}{\epsilon}-\gamma_E+ \log(4 \pi)$ one subtracts
$\Delta_m = \Delta -\log(m^2_M/\mu^2_{\DRbar})$, hence, avoiding large logarithms of the large scale $m_M$.
As studied in other works~\cite{decoup1}, this scheme minimizes higher order corrections
 when two very different scales are involved in a calculation 
of radiative corrections.
%Usually this choice is referred to in the literature as 'decoupling the large mass scale by hand' 
%(see e.g.\ \cite{decoup1,decoup2} and references therein).

Since we are interested in exploring the relevance of the new radiative corrections to the lightest
Higgs mass from the neutrino/sneutrino sector, we will present here our results in terms of the mass
difference with respect to the MSSM prediction. Consequently, we define,
\begin{equation}
\Delta m_h^{\mDRbar} := \Mh^{\nu/\Snu} - \Mh, 
\end{equation}
where $\Mh^{\nu/\Snu}$ denotes the pole for the 
light Higgs mass including the $\nu/\Snu$
corrections (i.e.\ in the MSSM-seesaw model), and $\Mh$ the corresponding pole in the MSSM, i.e
without the $\nu/\Snu$
corrections. Thus, for a given set of input parameters we first calculate $\Mh$ in the 
MSSM with the
help of \fh~\cite{mhcMSSMlong}, such that all relevant known 
higher-order corrections are included, and then we compute $\Delta m_h^{\mDRbar}$.
The full one-loop neutrino/sneutrino corrections to the self-energies,
$\hSi_{hh}^{\nu/\Snu}$, $\hSi_{HH}^{\nu/\Snu}$ and $\hSi_{hH}^{\nu/\Snu}$,
entering in the evaluation of $\Delta m_h^{\mDRbar}$ 
  have been evaluated with the help of 
\fa~\cite{feynarts} and  \fc~\cite{formcalc}.
\section*{Analytical and Numerical Results}
We have fully explored the dependence with all the new parameters in~\cite{Heinemeyer:2010eg} 
and found that the radiative corrections to the Higgs mass are mostly 
sensitive to $\mM$, $m_\nu$, $B_\nu$ and $\mR$. We focus here on $\mM$, $m_\nu$ and 
$\mR$ leading to the largest corrections and address the reader 
to~\cite{Heinemeyer:2010eg} for the complete study
in terms of all the parameters. 

First, in order to compare systematically our predictions of the neutrino/sneutrino
sector in the MSSM-seesaw 
with those in the MSSM, we have split the full one-loop neutrino/sneutrino
result into two parts:
\begin{equation}
\label{split}
\hSi(p^2)|_{\rm full}=\hSi(p^2)|_{\rm gauge}+\hSi(p^2)|_{\rm Yukawa}~,
\end{equation}  
where $\hSi(p^2)|_{\rm gauge}$ means the contributions from pure gauge 
interactions and they are obtained by switching off the Yukawa interactions,
i.e.\ by setting $\Ynu= 0$ 
(or equivalently $\mD=0$). The remaining part is named here 
$\hSi(p^2)|_{\rm Yukawa}$ and refers to the contributions that are only 
present if $\Ynu \neq 0$. In other words, this separation splits the full
result into the common part with the MSSM, given by $\hSi(p^2)|_{\rm gauge}$, 
and the new contributions due to the presence of Majorana neutrinos with 
non vanishing Yukawa interactions, given by 
$\hSi(p^2)|_{\rm Yukawa}$. Thus, by comparing the size of these two 
parts, within the allowed parameter space region, we will localize the 
areas where $\hSi(p^2)|_{\rm Yukawa} \gg \hSi(p^2)|_{\rm gauge}$, which  
will therefore indicate a significant departure from the MSSM result.   
%%%%%%%%%%%%%%%%%%%%%%%%%% F I G U R E %%%%%%%%%%%%%%%%%%%%%%%%%%%%%%%%%%%%%%%
 \begin{figure}[ht!]
   \begin{center} 
     \begin{tabular}{cc} \hspace*{-8mm}
  	\psfig{file=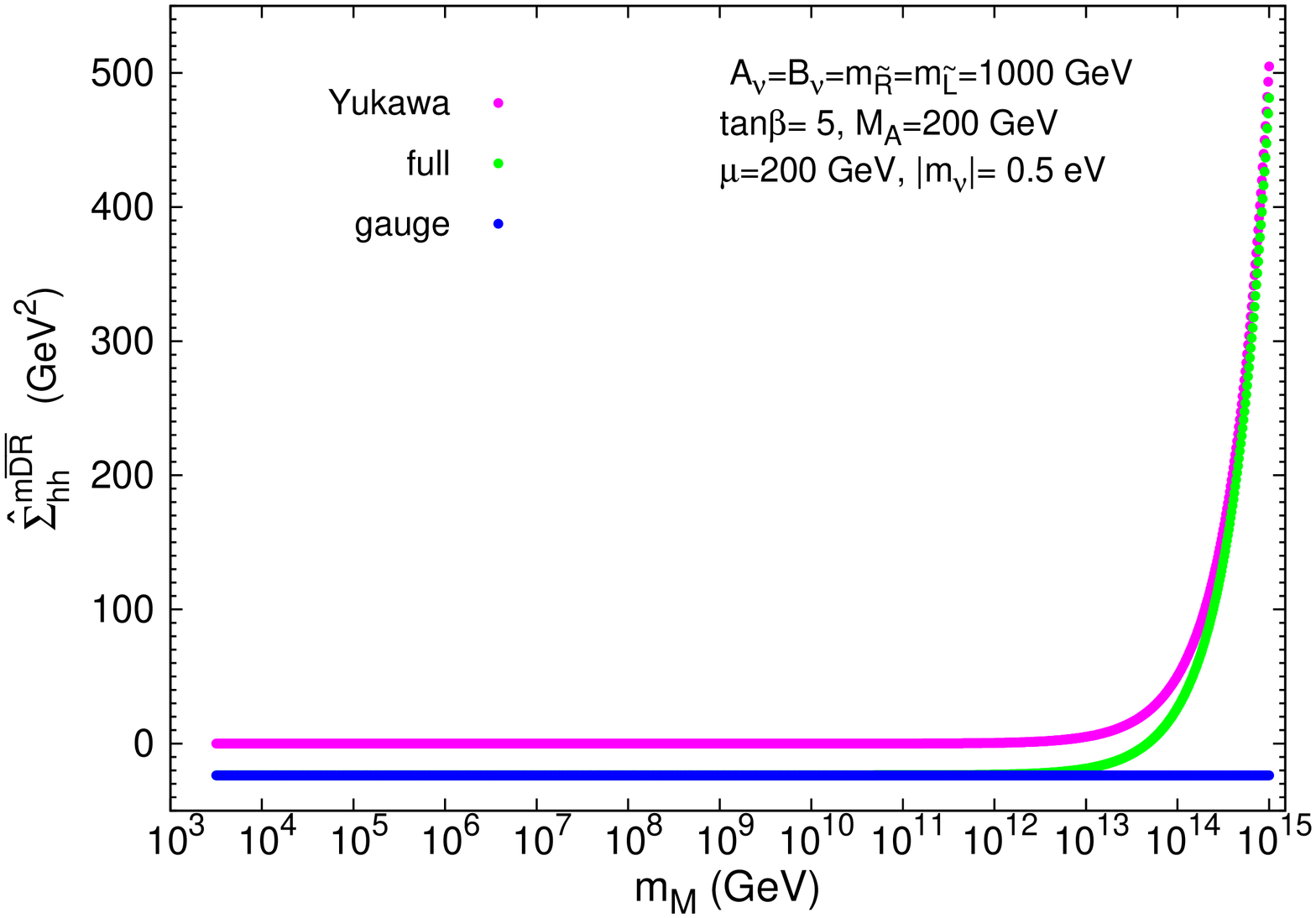,width=45mm,angle=270,clip=}  
  &
  	 \psfig{file=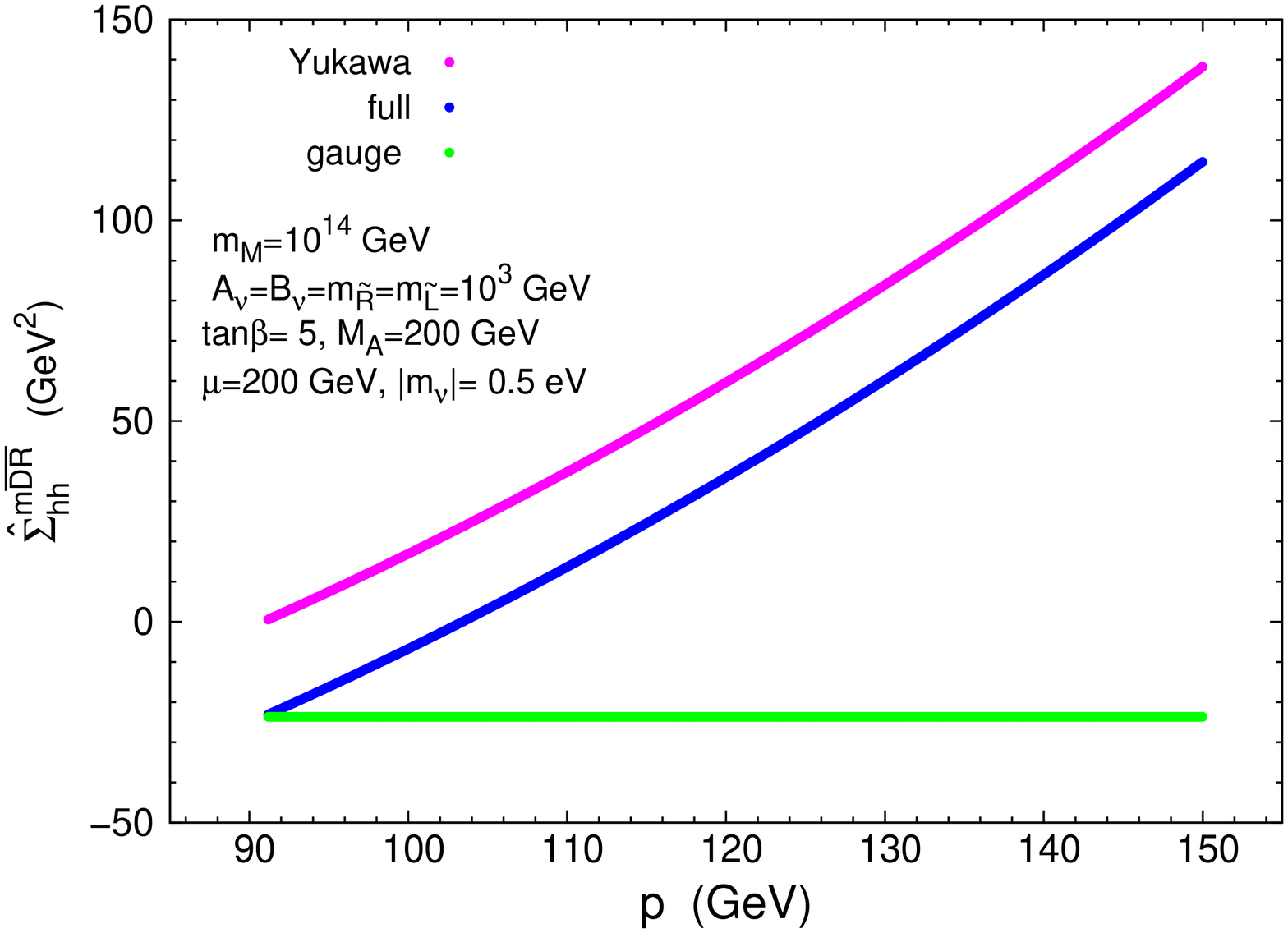,width=45mm,angle=270,clip=}  
      \end{tabular}
\caption{Left panel: Renormalized Higgs boson self-energy as a function of  
$\mM$. Here, $p= 116 \gev$. Right panel: Renormalized Higgs boson self-energy 
     as a function of the external momentum $p$. In both panels, the two 
     contributions from the
     gauge and Yukawa parts and the full
     result are shown separately.}
\label{fig:renSEversusmM}
   \end{center}
 \end{figure}
We see in Fig. \ref{fig:renSEversusmM} that the Yukawa part clearly dominates 
for large $\mM \sim  10^{14-15}$ GeV and that it shows a relevant external 
momentum dependence. Therefore, at large $\mM$, to keep just the Yukawa part is
a good approximation, but to neglect the momentum dependence or to set 
the external momentum to zero are certainly not.
In consequence, the alternative effective potential method will not provide a
realistic result for the radiative corrections to the Higgs mass.   

 Second, in order to understand in simple terms the analytical behavior of our full numerical results we have expanded 
the renormalized self-energies in
powers of the seesaw parameter $\xi=\mD/\mM$:
\begin{equation}
\hSi(p^2)=\left(\hSi(p^2)\right)_{\mD^0}+\left(\hSi(p^2)\right)_{\mD^2}+
\left(\hSi(p^2)\right)_{\mD^4} + \ldots ~.
\label{seesawser}
\end{equation}
The zeroth order of this expansion corresponds to the gauge contribution and it does not depend on $\mD$ or $\mM$. The rest
of the expansion corresponds to the Yukawa contribution.
The leading term of this Yukawa contribution is the ${\cal O}(m_D^2)$ term, because 
it is the only one not suppressed by the Majorana scale. 
In fact it goes as $Y_\nu^2M^2_{\rm EW}$, where  $M^2_{\rm EW}$ denotes generically the electroweak scales involved, 
concretely, $p^2$, $M_Z^2$ and $M_A^2$. In particular, the ${\cal O}(p^2m_D^2)$  terms of the renormalized self-energy,
 which turn out to be among the most relevant leading contributions, separated into the neutrino and sneutrino contributions,
 are the following: 
\begin{equation*}
\left.\hSi_{hh}^{\mDRbar}\right|_{\mD^2 p^2} \sim
\left.
\eqFeyn{%
\begin{picture}(80,50)
\DashLine(10,25)(25,25){3}
\Text(8,25)[r]{$h$}
\DashLine(55,25)(70,25){3}
\Text(72,25)[l]{$h$}
\ArrowArc(40,25)(15,0,180)
\Text(40,46)[b]{$\nu_L$}
\ArrowArc(40,25)(15,-180,0)
\Text(40,4)[t]{$\nu_R$}
\end{picture}
}
+
\eqFeyn{%
\begin{picture}(80,50)
\DashLine(10,25)(25,25){3}
\Text(8,25)[r]{$h$}
\DashLine(55,25)(70,25){3}
\Text(72,25)[l]{$h$}
\DashArrowArc(40,25)(15,0,180){3}
\Text(40,46)[b]{$\Snu_L$}
\DashArrowArc(40,25)(15,-180,0){3}
\Text(40,4)[t]{$\Snu_R$}
\end{picture}
}\right|_{\mD^2 p^2}
\end{equation*}
\begin{equation}
\hspace{-1.0cm}\sim \hspace{0.3cm}\frac{g^2 p^2 \mD^2 c^2_{\alpha}}{64 \pi^2  \MW^2 s^2_{\beta} } + \frac{g^2 p^2 \mD^2 c^2_{\alpha}}{64 \pi^2  \MW^2 s^2_{\beta} }.
\end{equation}
Notice that the above neutrino contributions come from the Yukawa interaction
 $g_{h\nu_L \nu_R}$, which is extremely suppressed in the Dirac case but can
be large in the present Majorana case.
 On the other hand, the above sneutrino contributions come from
the new couplings 
$g'_{h\Snu_L\Snu_R}$, which are not
present in the Dirac case. 
It is also interesting to remark  that these terms, being $\sim p^2$ are absent in both the effective potential  and the RGE approaches. 

An useful formula for the Higgs mass correction has then be derived, by keeping
just the dominant ${\cal O}(m_D^2)$ contribution to the Yukawa part of the
renormalized self-energy: 
\begin{align} 
\Delta  m_h^{\mDRbar} &\simeq  
 - \frac{\hSi_{hh}^{\nu/\Snu}(M_h^2)}{2M_h}  \approx
- \frac{\KL \hSi_{hh}^{\mDRbar}(M_h^2) \KR_{\mD^2}}{2 \Mh} \approx \non \\
& 
\frac{-g^2 \mD^2}{128  \pi^2\MW^2 \Mh\sin^2\be }  
 \KKL -2 M_A^2 \cos^2(\al-\be)\cos^2\be +2 \Mh^2 \cos^2\al\right.\nonumber \\
&\left.  - M_Z^2 \sin\be\sin(\al+\be)\left(2
\left(1+\cos^2\be\right)\cos\al-\sin2\beta\sin\al \right) \KKR.
\end{align} 
We have checked that this simple expression reproduces very accurately the full
numerical result, for the region of our interest with large $\mM \sim
10^{14-15}$ GeV.

\begin{figure}[ht!]
   \begin{center} 
     \begin{tabular}{c} \hspace*{-12mm}
  	\psfig{file=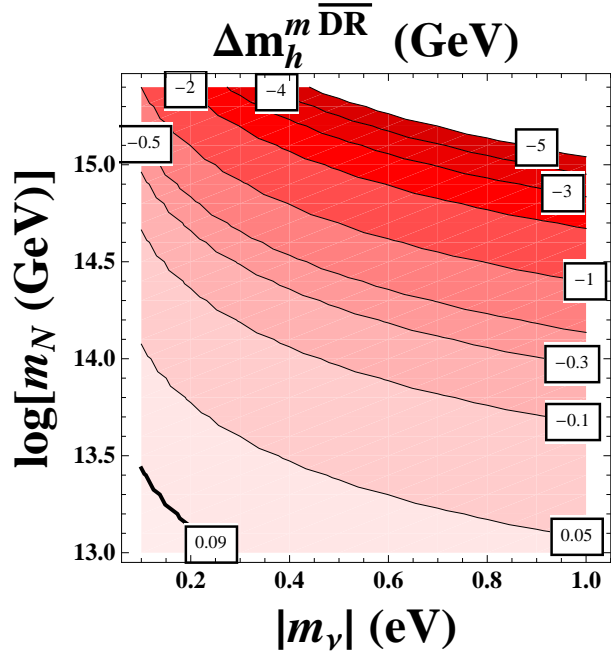,width=80mm,clip=}   
       \end{tabular}
     \caption{Contour-lines for the Higgs mass corrections from the
     neutrino/sneutrino sector as a function of the physical  
     Majorana neutrino masses, light $|\mnu|$ and heavy $m_N$. 
     The other parameters are 
     fixed to: $A_\nu=\Bnu=m_{\tilde L}=\mR=
     10^3 \gev$, $\tb=5$, $M_A=200 \gev$ and $\mu=200 \gev$.}  
   \label{masscontours1} 
   \end{center}
 \end{figure}

With respect to the numerical results, figure  \ref{masscontours1} exemplifies the main features of the extra Higgs mass corrections 
$\Delta m_h^{\mDRbar}$
due to neutrinos and sneutrino loops
 in terms of the two physical Majorana neutrino masses, $m_N$ and $\mnu$.
 For values of $m_N < 3 \times 10^{13}$ GeV and $|\mnu|< 0.1 -0.3$~eV
 the corrections to $M_h$ are positive and smaller than 0.1 GeV. In this region, the gauge contribution dominates. In fact,
 the wider black  contour line with fixed $\Delta m_h^{\mDRbar}=0.09$
coincides with the prediction for the case where just the gauge part in the
self-energies has been included. This means that 'the distance' of any other
contour-line with respect to this one represents the difference in the
radiative corrections respect to the MSSM prediction.

However, for larger values of  $m_N$ and/or $|\mnu|$ the Yukawa part dominates, and the radiative corrections
become negative and larger in absolute value, up to  values of -5 GeV in the right upper
 corner of Fig. \ref{masscontours1}. These corrections grow in modulus  proportionally to $\mM$ and $\mnu$, due to the fact
that the seesaw mechanism imposes a relation between the three masses  involved,
$\mD^2 = |m_{\nu}| m_N$.
%%%%%%%%%%%%%%%%%%%%%%%%%% F I G U R E %%%%%%%%%%%%%%%%%%%%%%%%%%%%%%%%%%%% 

\begin{figure}[h!]
   \begin{center} 
     \begin{tabular}{c} \hspace*{-12mm}
  	\psfig{file=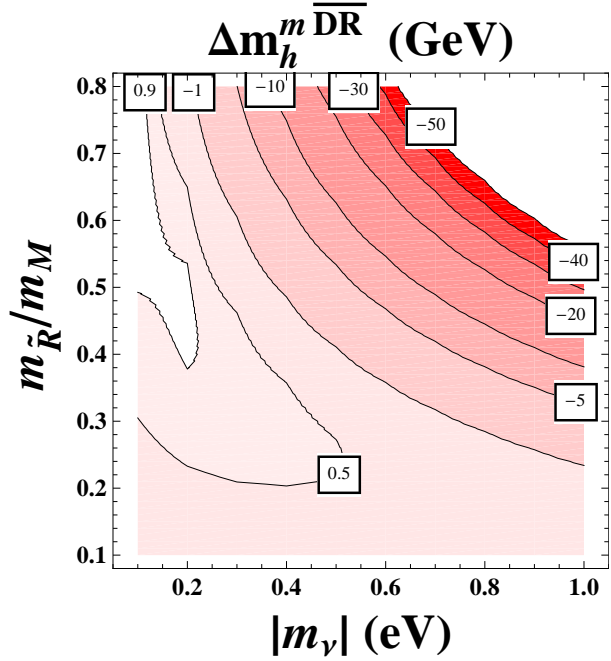,width=80mm,clip=}   
       \end{tabular}
     \caption{Contour-lines for the Higgs mass corrections from the
     neutrino/sneutrino sector as a function of the ratio 
     $\mR/\mM$ and the lightest Majorana neutrino mass $|\mnu|$. 
     The other parameters are fixed to: $\mM= 10^{14} \gev$, $A_\nu=\Bnu=m_{\tilde L}=
     10^3 \gev$, $\tb=5$, $M_A=200 \gev$ and $\mu=200 \gev$}  
   \label{masscontours_largemR} 
   \end{center}
 \end{figure}

Finally, we plot in Fig.~\ref{masscontours_largemR}, the contour-lines for fixed
$\Delta m_h^{\mDRbar}$ in the less conservative case where $\mR$ is
close to $\mM$. These are displayed as a function of $|\mnu|$ and the ratio
$\mR/\mM$. $\mM$ is fixed here to the reference value, $\mM=10^{14}$
GeV. For the interval studied here, we see again that the radiative corrections 
can be negative and as large as tens of GeV in the upper right corner of the
plot. For instance, $\Delta m_h^{\mDRbar}=-30 \gev$ for $\mM=10^{14}$
GeV, $|\mnu|= 0.6$ eV and $\mR/\mM= 0.7$.
%%%%%%%%%%%%%%%%%%%%%%%%%% F I G U R E %%%%%%%%%%%%%%%%%%%%%%%%%%%%%%%%%%%% 

%%%%%%%%%%%%%%%%%%%%%%%%%%%%%%%%%%%%%%%%%%%%%%%%%%%%%%%%%%%%%%%%%%%%%%%%%%%%

\section*{Conclusions}
We have used the Feynman diagrammatic approach for the calculation of the radiative corrections to
the lightest Higgs boson mass of the MSSM-seesaw.
This method  does not neglect the external momentum of the incoming and outgoing particles as it happens 
in the effective potential approach. We have performed a full calculation, obtaining not only the leading logarithmic terms 
as it would be the case in a RGE computation but also the finite terms, that
we have seen that can be sizable for heavy Majorana neutrinos ($10^{13}-10^{15} \gev$) and the lightest neutrino mass
within a range inspired by data
($0.1-1$~eV). For some regions of the MSSM-seesaw parameter space, the corrections to $\Mh$
are substantially larger  
than the anticipated LHC precision ($\sim 200 \mev$)~\cite{lhctdrs}.
Specifically they can be negative and up to -5 GeV if $\mM \sim 10^{15}$ GeV and
$|m_\nu|\sim 1$ eV and up to minus tens of GeV if in addition $\mR$ is also
large and of similar size to $\mM$.   
\section*{Acknowledgements}
M. Herrero wishes to thank the organizors of RADCOR2011 for this fruitful and
enjoyable meeting at the impresive site of Mamallapuram, India.

\end{document}